\documentclass[
superscriptaddress,
nofootinbib,
aps
]{revtex4-1}
\usepackage{graphicx}
\usepackage{epsfig}
\usepackage{amsmath}
\usepackage{amsfonts}
\usepackage{amssymb}
\usepackage{bm}
\usepackage[utf8]{inputenc}
\usepackage{natbib}
\usepackage{subfigure}
\usepackage{verbatim}
\usepackage{color}
\usepackage[colorlinks=true,citecolor=blue,urlcolor=blue,linkcolor=blue,menucolor=blue]{hyperref}
\hypersetup{breaklinks=true}
\usepackage{color}

\newcommand \be{\begin{eqnarray}}
\newcommand \ee{\end{eqnarray}}
\begin{document}
\begin{center}
\begin{Large}
{\bf Nuclear Response Functions with Realistic Interactions}\\
\bigskip
\bigskip
H. S. K\"ohler \\
\end{Large}
\footnote{e-mail:kohler@physics.arizona.edu}
{Physics Department, University of Arizona, Tucson, Arizona
85721,USA}\\ 

\end{center}

\date{\today}

\begin{abstract}
Linear density response functions are calculated for symmetric nuclear matter
of normal density by time-evolving two-time Green's functions in real
time. The feasability and convenience of this approach to this particular
problem has been shown in previous publications. Calculations are here 
improved by using more 'realistic' interactions derived from phase-shifts 
by inverse scattering.

Of particular interest is the effect of the strong correlations in the
nuclear medium on the response. This as well as the related energy
weighted sum rule, dependence on mean field and effective mass are some 
of the main objects of this investigation.

Comparisons are made with the collision-less limit, the HF+RPA method.
The importance of vertex corrections is demonstrated.

\end{abstract}

\section{Introduction}
Response functions, the response of a many-body system to an external
perturbation is instrumental in our understanding of the properties and
interactions involved in the excitations of the system.
In the study of nuclear systems these response functions are of particular
interest when it comes to calculate the mean free path and absorption of e.g.
neutrinos in a neutron gas \cite{mar04,iwa82}, a subject of interest in
astrophysical studies.\cite{red98,sed00}

They have been the subject of many  publications.
Nearly all reported
calculations use the "HF+RPA" method with Skyrme and/or Gogny effective forces.
\cite{gog77,gar92,ols04,mar05,mar06,mar08,sed10,gam11,pas12,pas14,pas15,pac16,nak17}
This method ignores the pre-existing correlations in the nuclear medium.
But nuclei are  strongly correlated many body systems.
It is however not trivial to include the effect of these correlations.
Simply dressing nucleon-propagators with self-energies leads to
inconsistencies.
Baym and Kadanoff \cite{bay62} showed that
appropriate vertex corrections
are also necessary to guarantee the preservation of the local continuity
equation for the particle density and current in the excited system.
This in turn implies the satisfaction of the important energy-weighted sum-rule.

These issues were investigated in detail in a previous work.\cite{hsk17}
A local interaction, independent of relative momentum, was used which 
allowed for a proof and test of  relations, such as the sum-rule. 
This interaction 
also made it possible to use an existing 2-time Kadanoff-Baym computer-code.
This work served to illustrate the importance of including correlations 
of the medium.

Although the properties of the potential
were adjusted to comply with known Landau parameters, it was still
deficient, e.g. being independent of relative momentum, a known important
property of effective interactions in nuclei.
Response calculations including the effect of in-medium correlations but with
a \it realistic \rm interaction is  called for.
Our choice of interaction is discussed below. (Section 2).  
Section 2.1 introduces our choice: Separable interactions constructed by
inverse scattering. In Section 2.2 these interactions are used in
Brueckner calculations .and in Section 2.3 in Green's
function calculations  of nuclear matter. Our linear response equations  are
shown in Section 3. with a discussion of the effective mass in section
3.1 Numerical results are shown in Section 4 with the HF+RPA in Sect 4.1
and correlations included in Section 4.2. A summary and some conclusions
are found in Section 5.

\section{NN Interactions}

The known NN-interaction has a short-ranged repulsive component with
high energy momentum representation. But the collisions in a nucleus 
are typically of low energy, of the order of the fermi-momentum. 
A major breakthrough in our understanding of nucleon interactions in a
nucleus is a realisation that these low-energy interactions can 
(with some caution) be represented by a low energy 'version' of the
interaction derived either by renormalising a high energy version 
as in $V_{low-k}$ or by EFT power-counting methods.

An important requirement of any realistic NN-potential model is
that it reproduces 'free', momentum dependent scattering phase-shifts.
A low energy version of the NN interaction  can then be defined  by a 
cutoff in momentum space with the requirement  that physical quantities,
such as the phase-shifts are reproduced up to this cut-off.
The practical impact of this  low-energy NN interaction is that is allows
for a perturbative calculation of nuclear properties \cite{hebeler,bogner}, 
as opposed to a typical Brueckner ladder summation to all orders.
Modern realistic low-energy potential  of this `type derived by EFT 
methods or $V_{low-k}$ are available.

Present computer (and programmming) limitations  prohibits  the 
detailed complexity of
these modern nucleon-nucleon interactions for response calculations.
It does however seem reasonable that  the NN-potential of choice should 
be realistic in the sense that it reproduces scattering data  and that 
it adequately reproduces the binding energy of nuclear matter
as well as mean field data such as effective mass etc to the extent that they
are known and affct the outcome of the calculation of interest.

In some previous publications on response functions we used a local Gaussian 
potential, used in earlier 2-time Kadanoff-Baym calculations. This choice 
was made partly because of the theory of response such as the energy  
weighted sum rule could be well documented within this frame work. Another 
reason for that choice was that the existing 2-time program was designed for 
local interactions only.\cite{hsk16}
Other authors used Gogny or Skyrme interactions for response calculations
typically by the HF+RPA method, e.g.  \cite{gog77,gar92}.

It is  however desirable to use a more realistic interaction, reelistic in the
sense defined above, while still allowing a reasonable computing effort.
A 2-body interaction that satisfies these requirements is  derived by a
purely phenomenological approach,  inverse scattering.

See following section for details.

\subsection{Separable NN interaction.}
For the response calculations shown below we are using non-local 
separable  potentials constructed by an inverse scattering method.
\cite{martin1,cha92,tabakin,kwo97,jisp16}. It 

Historically the first separable potential was constructed by Yamaguchi
\cite{yamaguchi1}. A  well-known attractive feature of  such a
potential is that it is easier to utilize in many-body calculations
since equations are simplified as for example
in the context of the Faddeev equations. That was for example
the reason for developping a
separable version of the Paris potetial.\cite{hai84}

One can in general construct an infinite
number of  NN interactions which are phase shift equivalent i.e. which fit
the on-shell properties of the scattering matrix, but may have 
different off-shell behaviors \cite{Bargmann}
Another attractive feature in addition to the one mentioned above 
is however that one can adjust
the off-shell components while keeping the on-shell intact by increasing the rank. 
Unlike the on-shell the off-shell is however not readily available
experimentally, other than indirectly from  deuteron data for 
example as in ref..\cite{kwo97}
One may of course also make adjustments to agree with data from some
realistic \it ab-initio \rm interaction, as in the afore-mentioned separable Paris
potential.

A potential derived by inverse scattering can be termed  realistic 
since it fits  the NN phase shifts at any given laboratory energy, although 
not termed $\textit{ ab-initio}$ in the sense that it does  not stem from 
an underlying theory of  strong interactions. 
Its authenticity is further supported by results of binding energy
calculations being (almost) identical to those of the Bonn-B potential,
in particular as regards the contribution from S-states.\cite{kwo97}
The triton binding energy as well as the n-D scattering length was also
well reproduced.\cite{hsk09}.

For reasons of simplicity in this first presentation of our method, we will 
include only the S-states (singlet and triplet) and neglect the tensor
coupling.

The separable potential will be  a function of the momentum cut-off 
$\Lambda$ up to which the phase shifts are fitted.
Below we show some results of second order (and Brueckner) calculations
that lead us to choose  $\Lambda=2$  fm$^-1$ for the response calculations.
This together with ignoring the coupling to $^3D_1$ states 
allows us to use rank one separable potentials for the $^1S$ and $^3S$ states
respectively.
This rank one potential has the simple form:

\begin{equation}
V_{\Lambda_{\rm cut}}(k,k^{\prime}) = \lambda \upsilon(k)
\upsilon(k^{\prime})
\end{equation} 
where $\lambda$ = $\pm$ 1 and
$\upsilon(k)$ is the numerical potential form factor which depends on
scattering phase-shift via the relation:

\begin{equation}
\upsilon^2(k) = -\lambda
\frac{(4\pi)^2}{k}\sin\delta(k)|D(k^2)|,
\end{equation}
As in previous works \cite{kwo97,hsk07,hsk09}  we choose to use the 
phase shifts $\delta(k)$ from ref. \cite{arndt}.  The function $D(k^2)$ is defined by:
\begin{equation}D(\omega) = \frac{\omega + E_{B}}{\omega} \exp \Big[
\frac{2}{\pi}  \int_{0}^{\infty} \frac{k^{\prime}
\delta(k^{\prime})}{\omega - k^{\prime 2}} dk^{\prime}  \Big],
\end{equation}
with the argument $\omega$ being in general complex and
the parameter $E_B$ standing for energy of the bound state in a specific
channel. In our case $E_B=0$ since the tensor force is not included and
none of the channels has a bound state. 
\begin{figure}
\begin{center}
\includegraphics[width=0.60\textwidth,angle=0]{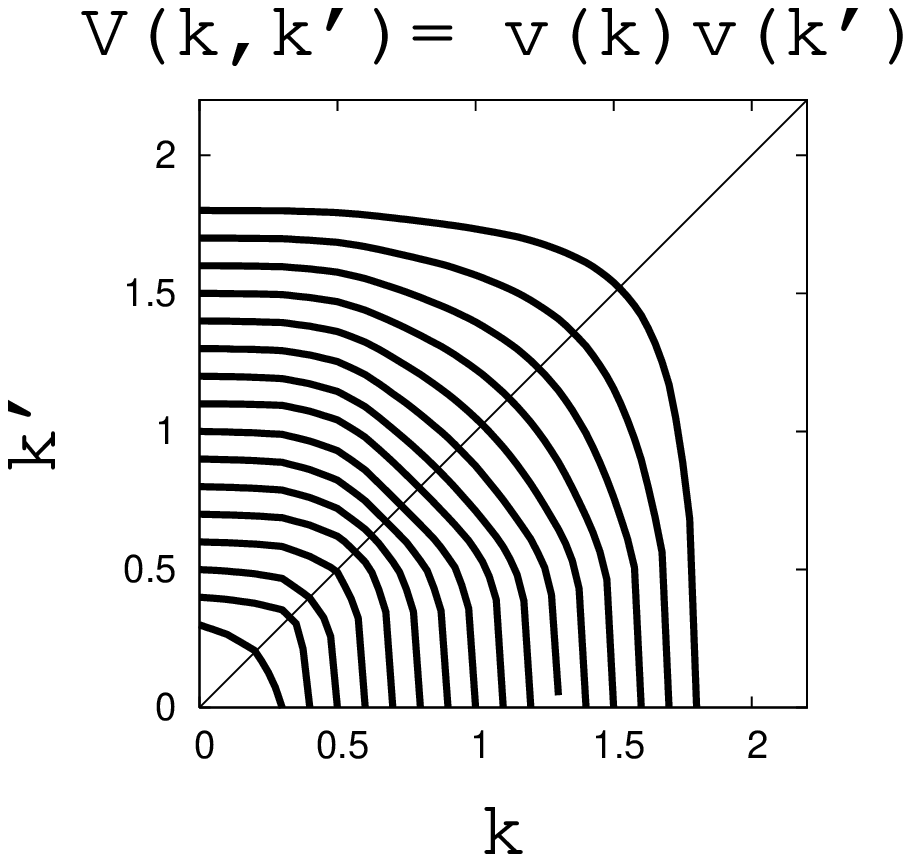}
\includegraphics[width=0.37\textwidth,angle=0]{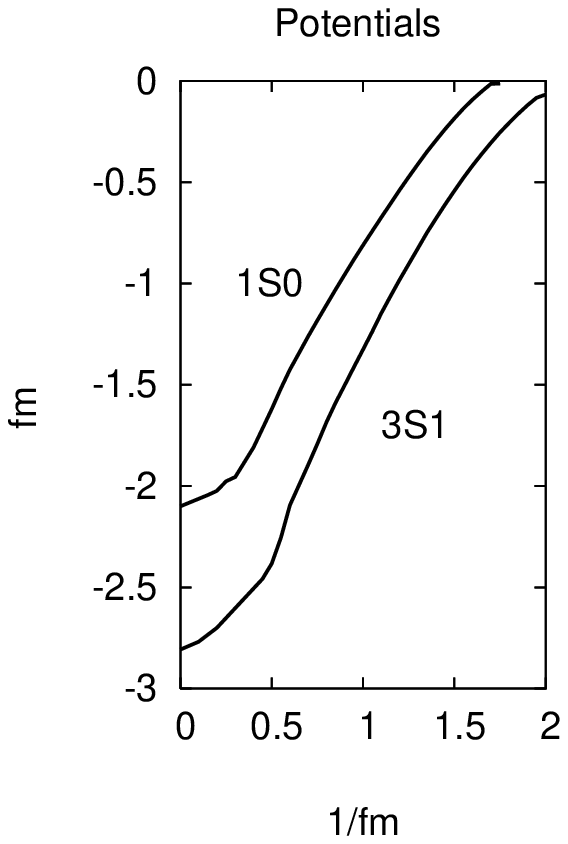}
\caption{On the left is shown a contourplot of the separable $^1S_0$
potential with cut-off $\Lambda=2$  fm$^{-1}$ that is used in
calculations shown below. On the right is shown 
the diagonal elements of both the $^1S_0$ and the $^3S_1$ potential.
}
\label{contour}
\end{center}
\end{figure}

A potential is of course not fully defined by fitting to some
phaseshifts. As pointed out above there are an infinite number of
soultions to that problem. But as also pointed out above, the
contribution to the binding enrgy of nuclear matter duplicates that of
the Bonn-B potential, which is a test of off-diagonal (off-shell)
components of the interaction. This is also exhibited by Figs below.
Fig. \ref{contour} shows a contourplot (left frame) of the
separable $^1S_0$ potential used in the calculations of response
functions shown below. The right frame shows the diagonal part
$V(k,k)=v(k)v(k)$ of the same potential. The potential is
calculated from eqs above with a cut-off $\Lambda=2$ fm$^{-1}$. These results are comparable with the  similar display in ref. 
\cite{hebeler} (Figs 3 and 17) of the $V_{low-k}$  interaction. 
Both interactions are momentum-dependent i.e. non-local. 
The overlap between the two, the $V_{low-k}$ and the separable is
compelling. Note however that this observation refers only to the
$^1S_0$ state.

The similarity can be understood by the following :
The $V_{low-k}$ interaction is analogous to that defined by the separation
method due to Moszkowski and Scott (MS) \cite{mos60} as shown by Holt and Brown
\cite{hol04}. The formal difference is that the former as well as our separable 
potential is defined by a cut-off in momentum-space while the latter  
by a cut-off in coordinate-space.
We point out that the S-state component of a local interaction is non-local 
although not (necessarily) separable. (See e.g. ref. \cite{hsk65}).
The MS-potential is zero within a separation distance  $d \sim 1$ fm, and thus
represented by a hollow shell. In the limit of approximating this potential by a
'hard' shell at some distance $d_{eff}$ this potential is local but the
S-state component  of this potential is  separable. \cite{mospriv}
A Gaussian separable potential was used in ref. \cite{hsk62} as an
approximation of the MS-potential with (almost) identical overlap.
It was there used in a Brueckner calculation of $^{16}O$ in a Harmonic
Oscillator basis.

\subsection{Nuclear Matter by Brueckner theory.}
\begin{figure}
\begin{center}
\includegraphics[width=0.52\textwidth,angle=0]{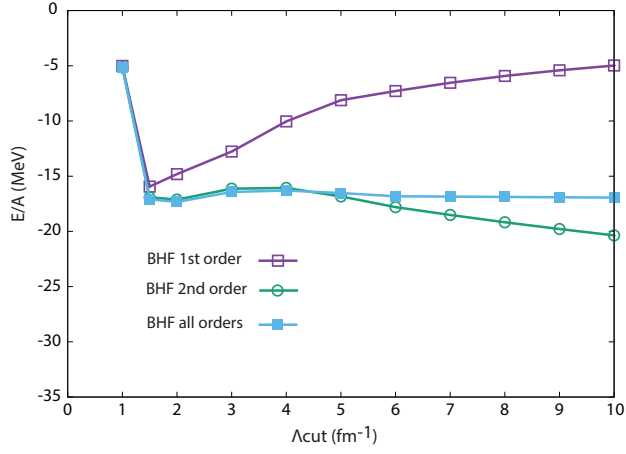}

\caption{
Nuclear matter energy per particle as a function of momentum cut-off
$\Lambda_{\rm{cut}}$  for indicated  orders of the interaction,
including $^1S_0$ and $^3S_1$ channels only.}
\label{Brueckner}
\end{center}
\end{figure}

Using the formalism above  we construct separable potentials for momentum 
cut-offs  ranging from $\Lambda = 1$ fm$^{-1}$ up to $\Lambda = 10$ fm$^{-1}$ ( including  only the $^1$S$_0$ and $^3$S$_1$ channels ) 
and we perform 
Brueckner calculations for symmetric infinte nuclear matter.  Table I. 
shows  total energys as a function of $\Lambda$  in first, second and 
all orders of the interaction.  Fig. \ref{Brueckner}  displays the same data.
The fermi-momentum is $k_F=1.25$ fm$^{-1}$.
  \begin{table}[h!]\caption{Nuclear matter  energies in MeV at several
  orders as a function of the parameter $\Lambda$.} 
\centerline{
  \begin{tabular}[t]{|c |c |c |c |}
\hline
  $ \Lambda$
  (fm$^{-1}$) & 1$^{\rm st}$ order & 2$^{\rm nd}$ order & all orders \\
  \hline 1   &  -5.02      &   -5.14     &  -5.16   \\
  \hline 1.5  &   -15.94     &     -16.91   &   -17.09 \\
  \hline2   &  -14.81      &      -17.12  &    -17.33\\
  \hline3   &  -12.77      &   -16.13     &    -16.43\\
  \hline4   &    -10.04    &    -16.06    &    -16.30\\
  \hline5   &   -8.12     &   -16.83     &    -16.53\\
  \hline6   &   -7.29     &  -17.80      &    -16.82\\
  \hline7   &    -6.54    &    -18.51    &    -16.85\\
  \hline8   &     -5.92   &    -19.17    &    -16.88\\
  \hline9   &     -5.41   &  -19.78      &    -16.91\\
  \hline 10  &    -4.98    &   -20.36     &    -16.94\\
  \hline 
  \end{tabular}
 } 
  
  \label{tab:1}
  \end{table}
One sees that the second order  calculation produces results almost 
identical with the all orders calculation for  cutoffs $\Lambda$ 
ranging from $\sim$ 1 fm$^{-1}$ up to 5 fm $^{-1}$. We conclude that 
the separable interaction is  soft enough that even at cutoffs 
as large as 5 fm$^{-1}$ the second order calculation is a good 
approximation. 
Only the $^1$S$_0$ and the $^3$S$_1$ partial waves are included
and  the tensor force is also neglected.  Previous results \cite{kwo97} show 
that  contributions of higher angular momentum states almost 
cancel out; the main contributions  come from the S-waves. 
It is of course well-known that  including the tensor force is vital for the
saturation of nuclear matter. And so is the short-ranged correlations that are
not included when the cut-off is less than $\sim$ 2-3 fm$^{-1}$.\cite{hsk07}

The main purpose here is however to establish the
usefullness of the low-energy version of the separable potentials
(calculated by inverse-scattering) for response-calculations at
\it normal \rm nuclear matter density.
The effects of the tensor interaction in the calculation of response functions
within the two-time approach will be investigated in subsequent publications.

The results above show that with the S-states to second order together 
with a cut-off $\Lambda=$ 2 fm$^{-1}$ gives a binding energy of 17.12 MeV/A 
compared to $17.33$ MeV/A in an all order  summation. 
The 2 fm$^{-1}$ cut-off allows us to use  a rank one
separable potential.\cite{kwo97} This will be the choice of interaction in the
calculations to follow below.
Correlations in the response calculations will be included by second 
order self-energies.

\subsection{Nuclear Matter with 2-time Green's functions.}
The calculation of response functions follows the methods used in
earlier work, time evolving Green's functions by Kadanoff-Baym
equations. \cite{nhk00,hsk16}  The Green's functions are separated
into a spatially homogeneous part $G_{00}(t,t')$ and a linear 
response part $G_{10}(t,t')$.  Green's functions $G_{00}(t=0,t'=0)$ 
are constructed for an uncorrelated fermi distribution of specified 
density and temperature. The $G_{00}$ functions are then time-evolved
(for typically 10  fm/c)  with the chosen  selfenergies until
fully correlated.  Selfenergies are calculated to second order with the 
separable $^1S$ and $^3S$ interactions specified above.

The KB-equations for the propagation of these 
$G_{00}$ functions as well as numerical methods for solution  has 
already been shown in  previous works (e.g. \cite{hsk99}), but included
below for completeness.

(summation over $m=0,1$ and  integrations over $\bar{t}$ from $-\infty$
to $+\infty$ is implied):
\begin{eqnarray}
(i \frac{\partial}{\partial t}-\frac{p^{2}}{2m}-\Sigma_{00}^{HF}(p,t))
G_{00}^{^{>}_{<}}({\bf p},t,t')
=&(\Sigma_{00}^{>}({\bf p},t,\bar{t})-
\Sigma_{00}^{<}({\bf p},t,\bar{t}))G_{00}^{^{>}_{<}}({\bf p},\bar{t},t')-\nonumber \\
&\Sigma_{00}^{^{>}_{<}}({\bf p},t,\bar{t})
(G_{00}^{>}({\bf p},\bar{t},t')-G_{00}^{<}({\bf p},\bar{t},t'))
\label{eq1}
\end{eqnarray}
\begin{eqnarray}
(-i \frac{\partial}{\partial t'}-\frac{p^{2}}{2m}-\Sigma_{00}^{HF}(p,t'))
G_{00}^{^{>}_{<}}({\bf p},t,t')
=&(G_{00}^{>}({\bf p},t,\bar{t})-
G_{00}^{<}({\bf p},t,\bar{t}))\Sigma_{00}^{^{>}_{<}}({\bf p},\bar{t},t')-\nonumber \\
&G_{00}^{^{>}_{<}}({\bf p},t,\bar{t})
(\Sigma_{00}^{>}({\bf p},\bar{t},t')-\Sigma_{00}^{<}({\bf p},\bar{t},t'))1
\label{eq1a}
\end{eqnarray}

The past (known) versions of the two-time code limits the calculation of
self-energies $\Sigma^{^{>}_{<}}$ to the use of an interaction that
is local (in coordinate space), i.e. momentum independent.  A new version
of the KB-code has now been developped for the separable potentials
with the $|Sigma_{00}$  selfenergies given by:
\begin{equation}
\Sigma_{00}^{HF}({\bf p},t)=
i\sum_{\bf p',j}G^{<}_{00}({\bf p-p'},t,t)\lambda \upsilon_j^2(p')
\label{HF00}
\end{equation}
and
\begin{eqnarray}
\Sigma_{00}^{^{<}_{>}}({\bf p},t,t')=
i\sum_{\bf p_1,p_2,j}G_{00}^{^{<}_{>}}({\bf p_1},t,t')G_{00}^{^{>}_{<}}({\bf
p_2},t,t')G_{00}^{^{<}_{>}}({\bf p+p_2-p_1},t,t')\times \nonumber \\
\upsilon_j^2(2{\bf p_1}-{\bf p-p_2}) \upsilon_j^2({\bf p-p_2})
\label{S00}
\end{eqnarray}
where the index $j=1,2$ refers to the two $S$-states.

A diagrammatic representation of the self-energy $\Sigma_{00}$  is shown in 
Fig.  \ref{2Born}.

\begin{figure}
\begin{center}

\includegraphics[width=0.52\textwidth,angle=-90]{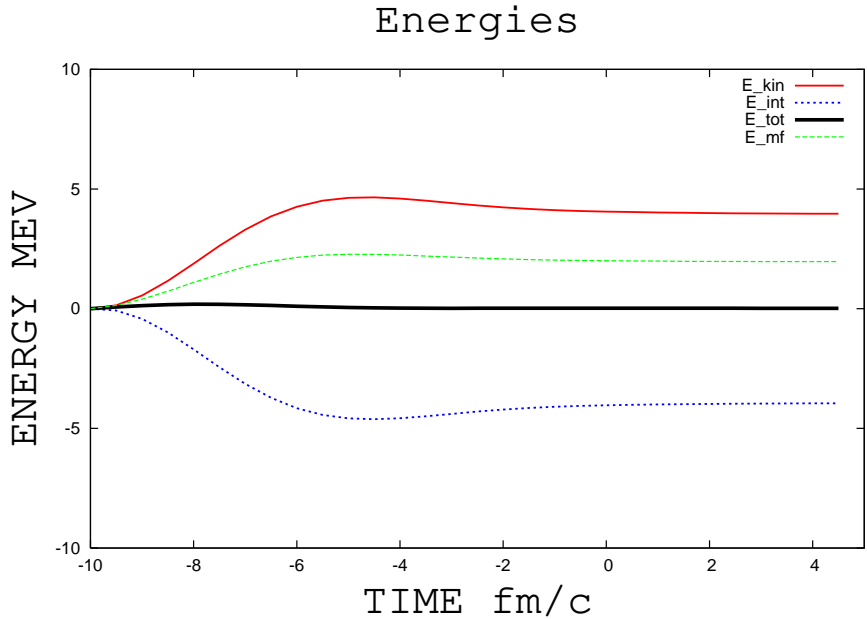}

\caption{
Energies as a function of time starting at t=-10 fm/c. 
Results shown are from top to bottom:
kinetic (red online),mean field (green online),  total (black online)
and potential (blue online)  energies. Energies are shifted  to 
 energy= zero at time $t=-10$. See text for further details.
}
\label{ener}
\end{center}
\end{figure}

Fig. \ref{ener} shows the separate energies, (kinetic, potential and total)
as a function of time in a calculation with the separable $^1S_0$ and 
$^3S_1$ interactions defined  above. The cutoff $\Lambda=2$fm$^{-1}$. 

The initial state  is here a zero-temperature fermi-distribution, 
uncorrelated.  The fermimomentum in this and in results below
are for symmetric nuclear matter with $k_f=1.25$ fm$^{-1}$.
The equations are time-stepped until system is fully correlated.  
The time-scale starts for convenience at $t=-10$ fm/c with the system 
considered fully correlated at $t=0$ with a correlation time 
$t_c=10$ fm/c.\cite{hsk01}  The external perturbation is applied 
as a pulse centered at $t=0$.  
All  energies are shifted to zero at $t=-10$. This is to better show 
the change in energies from the uncorrelated to the correlated state.
The total energy (kinetic+potential) is  constant in time,  
which is the result of 
conserving approximations for the selfenergies. \cite{bay62}
The interaction (potential) energy $E_{pot}$ includes both the 
mean field $E_{mf}$ and a 'correlation' energy $E_{corr}$.  
The initial  ($t=-10$) kinetic (same as total) and mean field energies are 
$19.4$ and $-37.4$ MeV respectively. 

At the end of the run the
kinetic,interaction and mean-field energies have changed by $+4.0a$,
$-4.0$ and $+2.0$ MeV respectively and the correlation energy
(the difference between the interaction and mean-field energies) 
$E_{corr}$ = $6.0$   MeV.
It might seem that this energy should be comparable with the
second order  contribution  in Section 2,2 , the difference between 
the first and second order results found in Table 1, which is seen to 
be $2.31$ MeV,, a difference of almost a factor of three. 
There are several reasons for this apparent "discrepancy". One is the
effect of the mean field, which is a consequence of the redistribution in
momentum-space shown below in Fig. \ref{dens}. Neglecting the mean 
field in each of the two
calculations the Brueckner gives $-4.1$ MeV for the second order Born 
contribution while the KB gives $-8.9$ MeV i.e. a factor of $\sim 2$.
A factor of exactly $2$ was already demonstrated to be the exact value
to be expected in the Levinson (and the extended quasiparticle)
approximation of the collision term.\cite{hsk01,hsk92a}. 
It is associated with the increase in kinetic energy (see Figs \ref{ener} and
\ref{dens}) with the KB method.

Fig. \ref{dens} shows the correlated distribution in momentum space. 
It is seen to compare reasonably well with the many previously published  
results at the fermi-surface.  (e.g. ref.\cite{hsk16}). The depletion of
interior states is however appreciably less. This is because of cut-off
of the large momenta (short ranged) as well as the neglect of the
tensor-componenet in these calculations.
\begin{figure}
\begin{center}
\includegraphics[width=0.52\textwidth,angle=0]{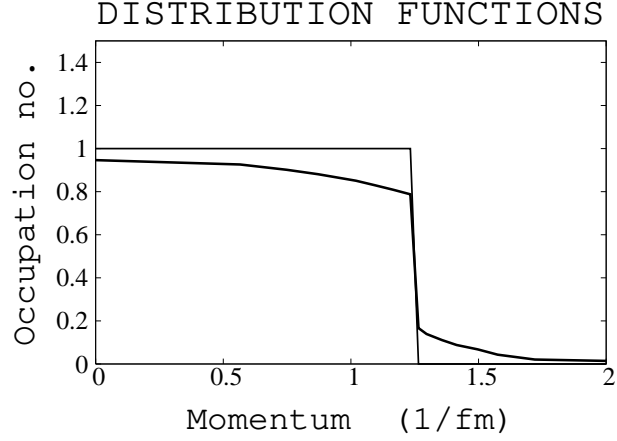}

\caption{
Shown here is the  zero temperature Fermi-distribution (top curve) 
and the correlated distribution from the second-order selfenergy calculations.
}
\label{dens}
\end{center}
\end{figure}

\section{Linear response with the separable NN-potential.}
The formalism associated with the calculation of the response-function
using the 2-time KB- method has been shown in previous works
\cite{nhk00,hsk16}. 

Eqs (\ref{eq1}) and (\ref{eq1a}) showed the time-evolution of the Green's
functions $G^{^{>}_{<}}_{00}$ for the unperturbed nuclear system.
At a correlation time $t=t_c$, ($t=0$ in Fig.\ref{ener}) this  system 
is 'hit' by an external potential $U({\bf q},t) =U_0(t)\delta_{{\bf q,q_{0}}}$
that results in collective excitations. These excitations are contained in
Green's functions $G_{10}^{^{>}_{<}}$, obeying the equations
(summation over $m=0,1$ and  integrations over $\bar{t}$ from $-\infty$
to $+\infty$ is implied):

\begin{eqnarray}
\left(i\hbar \frac{\partial}{\partial t}-\epsilon_{{\bf k+q_0}} \right )
G_{10}^{^{>}_{<}}({\bf k}tt')=U_0(t)G_{00}^{^{>}_{<}}({\bf k}tt')
+\Sigma_{1m}^{HF}({\bf k}t)G_{m0}^{^{>}_{<}}({\bf k}tt')
\nonumber \\
+\Sigma_{1m}^{R}({\bf k}t\bar{t})G_{m0}^{^{>}_{<}}({\bf k}\bar{t}t')+
\Sigma_{1m}^{^{>}_{<}}({\bf k}t\bar{t})G_{m0}^{A}({\bf k}\bar{t}t')
\label{eq01a}
\end{eqnarray}

and

\begin{eqnarray}
\left(-i\hbar \frac{\partial}{\partial t'}-\epsilon_{{\bf k}} \right )
G_{10}^{^{>}_{<}}({\bf k}tt')=U_0(t')G_{11}^{^{>}_{<}}({\bf k}tt')
+G_{1m}^{^{>}_{<}}({\bf k}tt')\Sigma_{m0}^{HF}({\bf k}t')
\nonumber \\
+G_{1m}^{R}({\bf k}t\bar{t'})\Sigma_{m0}^{^{>}_{<}}({\bf k}\bar{t}t')
+G_{1m}^{^{>}_{<}}({\bf k}t\bar{t'})\Sigma_{m0}^{A}({\bf k}\bar{t}t')
\label{eq01b}
\end{eqnarray}

\begin{figure}
\begin{center}
\includegraphics[width=0.52\textwidth,angle=0]{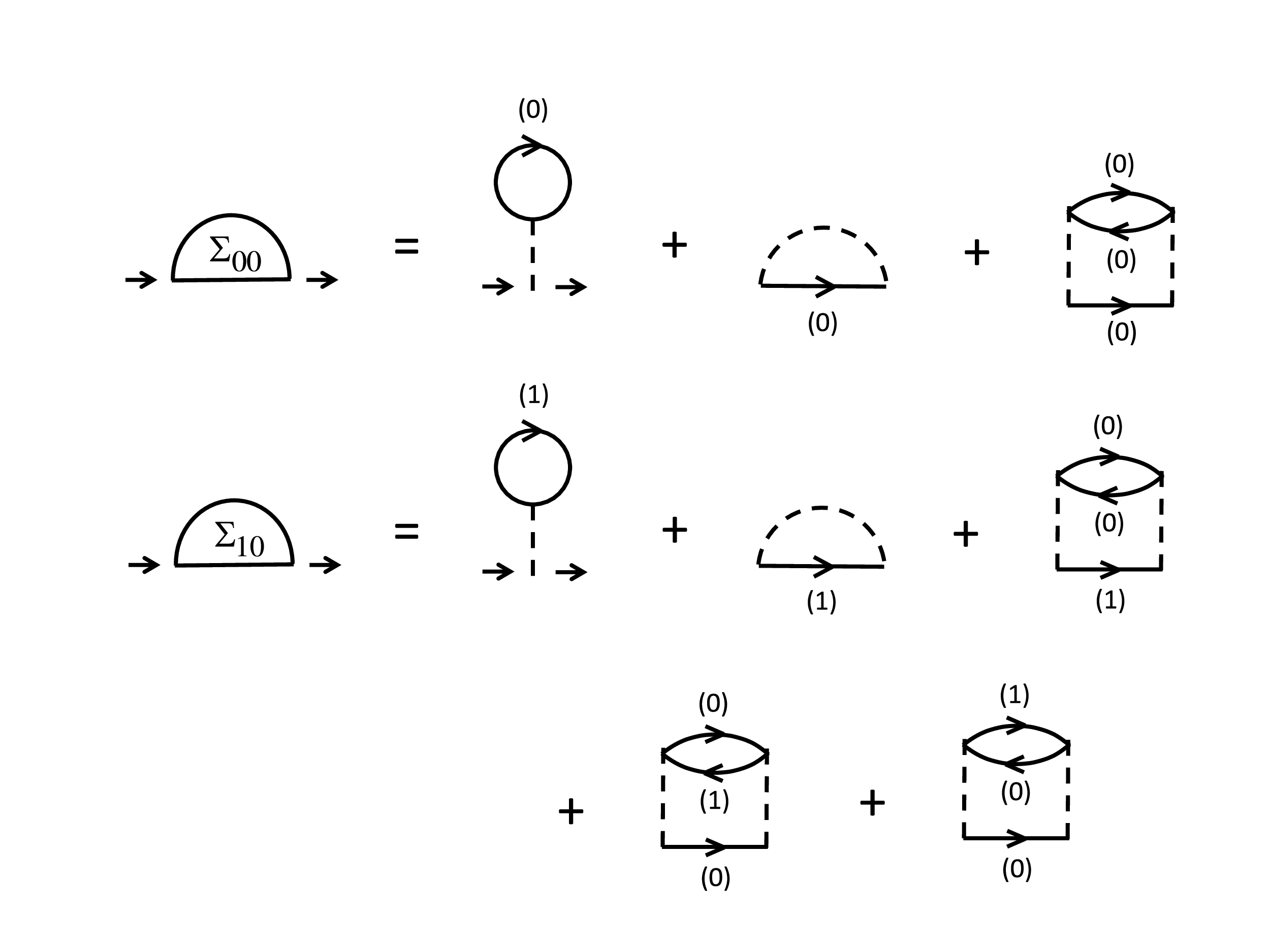}
\caption{
Diagrams representing contributions to the self energy in the
Kadanoff-Baym equations to zeroth order 
( $\Sigma_{00}$ ) and 
first order ( $\Sigma_{10}$ ) in the external perturbation $U$. 
The solid (dashed) lines are correlated Green's functions 
NN interactions). The number near each Green's function line 
gives the order in $U$ of that line.
}
\label{2Born}
\end{center}
\end{figure}

A diagrammatic representation of the self-energies is shown in Fig.
\ref{2Born}. 
(See also ref. \cite{hsk16} for a more complete exhibition.)
We distinguish between  three contributions to the self-energy
$\Sigma_{10}^{^{<}_{>}}$ corresponding to the three 
second order diagrams shown in Fig \ref{2Born} and  write
$$\Sigma_{10}^{^{<}_{>}}({\bf p},t,t')=
\sum_{n=1,3} \Sigma_{(n)}^{^{<}_{>}}({\bf p},t,t')$$
with
\begin{eqnarray}
\Sigma_{(1)}^{^{<}_{>}}({\bf p},t,t')=
i\sum_{\bf p_1,p_2}G_{10}^{^{<}_{>}}({\bf p_1},t,t')G_{00}^{^{>}_{<}}({\bf
p_2},t,t')G_{00}^{^{<}_{>}}({\bf p+p_2-p_1},t,t')\times \nonumber \\
\upsilon^2(2{\bf p_1}-{\bf p-p_2}) \upsilon^2({\bf p-p_2})
\label{S1}
\end{eqnarray}

\begin{eqnarray}
\Sigma_{(2)}^{^{<}_{>}}({\bf p},t,t')=
i\sum_{\bf p_1,p_2}G_{00}^{^{<}_{>}}({\bf p_1},t,t')G_{00}^{^{>}_{<}}({\bf
p_2},t,t')G_{10}^{^{<}_{>}}({\bf p+p_2-p_1},t,t')\times \nonumber \\
\upsilon(2{\bf p_1}-{\bf p-p_2})\upsilon(2{\bf p_1}-{\bf p-p_2+q_0})
\upsilon({\bf p-p_2})\upsilon({\bf p-p_2+q_0})
\label{S2}
\end{eqnarray}

\begin{eqnarray}
\Sigma_{(3)}^{^{<}_{>}}({\bf p},t,t')=
i\sum_{\bf p_1,p_2}G_{00}^{^{<}_{>}}({\bf p_1},t,t')G_{10}^{^{>}_{<}}({\bf
p_2-q_0},t,t')G_{00}^{^{<}_{>}}({\bf p+p_2-p_1},t,t')\times \nonumber \\
\upsilon^2(2{\bf p_1}-{\bf p-p_2}) \upsilon^2({\bf p-p_2})
\label{S3}
\end{eqnarray}
where 

\begin{equation}
G^{^{>}_{<}}_{10}({\bf k},t,t')\equiv
 G^{^{>}_{<}}_{10}({\bf k+ q_0},t;{\bf k},t')
\label{eq01c}
\end{equation}

The retarded and advanced parts above are given by

\begin{eqnarray}
\Sigma_{10}^{R/A}({\bf p},t,t')=\pm \theta(\pm
(t-t')[\Sigma_{10}^{>}({\bf
p},t,t')-\Sigma_{10}^{<}({\bf p},t,t')]
\label{eq5a}
\end{eqnarray}

Results of calculations are shown below following a discussion of the effective
mass.

\subsection{Effective mass}
The effective mass plays a very important role in the theory of
response-functions. It determines a mean  energy of the excitation as is
evident from the energyy weighted sum-rule: It will be seen below
that it also affects the width of the response function.
\begin{equation}
\int \omega S(\omega,q_0)d \omega=\frac{q_{0}^{2}}{2m^*}
\label{sum}
\end{equation}
We are in particular interested in the response in the long wave-length
limit with excitations close to the fermi-surface. The effect of the
external perturbation will depend on the
energy-spectrum $e(k)$ out of which the particles are excited,
conveniently expressed in terms of the effective mass
$m^*$  with ($\hbar=1$) $$e(k)=\frac{k^2}{2m}+U(k)=U(0)+\frac{k^2}{2m^*(k)}$$

The effective mass will, as indicated, in general be a function of $k$
with $m^*(k_F)$ being the effective mass of interest here.
It has been the subject of many calculations and discussions since
early works on the Landau theory and nuclear many body problem in general.
(see e.g.  \cite{jeu76}).  Of particular interest for our present work is
that of B\"ackman \cite{bac68} and Sj\"oberg \cite{sjo73}, related to
Landau theory.
In the calculations presented below we are defining an 'effective'
effective mass $m^{**}$ from inverting eq. \ref{sum}.
\begin{equation}
m^{**}=\frac{q^{2}_{0}}{2\int \omega S(\omega,q_0)d \omega}
\label{summ}
\end{equation}
This relies on the fact that our equations do satisfy the energy
sum-rule. This was tested and verified by replacing the Hartee- Fock
field $\Sigma_{00}^{HF}$ below in  eqs (\ref{eq01a}) and (\ref{eq01b}) by an
effective mass approximation. 

The definition of  $U(k)$ or equivalently $e(k)$ ,  is evidently of 
utmost importance. We are here concerned with excitations due to  
an external perturbation
and the definition relevant for the present work is then
\begin{equation}
e(k)=\frac{dE}{dn_{k}}
\label{m^*}
\end{equation}
i.e. the removal energy.
In Brueckner theory this would include terms to first order in the
Brueckner $K$-matrix as well as the higher order rearrangement terms.
There are numerous  publication and discussions in the literature on
this subject matter.  The third order term $U^{(3)}$ is related to 
the depletion factor $\kappa$ by $$U^{(3)}=-\kappa U^{(1)}.$$
K\"ohler and Moszkowski \cite{hsk07} evaluated contributions to this 
depletion factor for eight of the most important spin-isospin states 
for our separable potential with $1.6<\Lambda<9.8$ fm$^{-1}$. The results 
showed a strong dependence  on the cut-off $\Lambda$.
For the largest cut-off considered, $\Lambda=9.8$ fm$^{-1}$ they 
found $\kappa=.175$
for a density with $k_F=1.35$ fm$^{-1}$. For $\Lambda=2.6$ fm$^{-1}$ (the
closest to or chosen value $\Lambda=2.0$) they found $\kappa=0.124$.
The largest contribution was for the coupled $^3S_1$-$^3D_1$ states 
which we do not include at present.  It does however seem appropriate 
to here adopt the value $\kappa=.17$.

The second order term $U^{(2)}$ stems from the change in Pauli-blocking upon
removal of a nucleon. The significance  of this term was discussed early on
by Brueckner et al \cite{bru58} and later in refs. \cite{hsk65,sar80}
It is strongly momentum-dependent and thus quite important as regards 
the effective mass.  An increase of the effective mass near the fermisurface 
by $\sim 0.15$ compared to that for deeper states is expected.\cite{jeu76}

The modification (increase) of the effective mass from that given by a first 
order mean field calculation is of importance for the response calculations.
While our  first order result yields $m^*\sim 0.6$ the rearrangement terms
increases it to $m^*\sim 0.8-0.9$.  This should be compared with the Landau 
value close to $m^*=1.$, validated by experimental evidence\cite{bro63}. We
return to the question of the effective mass in the Results section below.

\section{Numerical results}
The formalism presented above is applied  to calculating response
functions of symmetric nuclear matter at normal density. 
To fully appreciate the importance of the various self-energies etc we
also show some results of approximations.

\subsection{HF+RPA}
\begin{figure}
\begin{center}
\includegraphics[width=0.52\textwidth,angle=0]{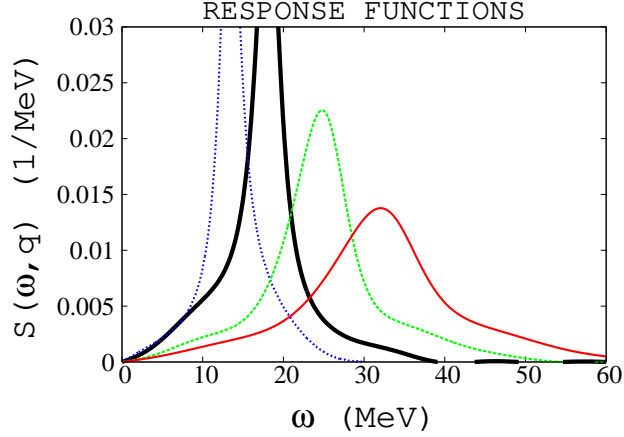}

\vspace{.2in}
\caption{
Response functions with mean-field only; no correlations i.e. by the HF+RPA
method.  From left to right: The first line (blue online) is with the
HF mean field $\Sigma_{00}^{HF}$ set to zero i.e. with
$m^*=m^{**}=1.$
The solid black line is with the HF-field 
calculated with the $^1S$ and $^3S$ separable potentials. It relates 
to an effective mass $m^{**}=0.66$. (See text regarding definition of 
$m^{**}$.) The next, broken line (green online), shows the result with 
the mean field replaced by an effective mass $m^*=0.5$ and the last 
curve, broken line (blue online), with an effective mass $m^*=0.4$. 
It is seen that there is not only  a shift in energy but also a
broadening of the response function when theffective mass decreases.
All results are with $q_0=0.4$ fm$^{-1}$. 
}
\label{HF}
\end{center}
\end{figure}

Most published reports on response functions  use  the HF+RPA
method. This implies neglecting all effects of
correlations, i.e. all  second order self-energies in Fig. \ref{2Born}.
while maintaining the mean fields.
Our results are shown in Fig. \ref{HF}, all with $q_0=0.4$  fm$^{-1}$.
The left curve shows the result with the mean field calculated
selfconsistently. The 'effective' effective  mass $m^{**}$ is here 
obtained from  eq.  (\ref{summ}) . The result is $m^{**}=0.66$.

In the two other results the mean field is, as indicated, replaced by an
effective mass approximation, which allows us to test and verify  the
energy sum rule. It also shows the importance of the
effective mass as it affects the response, a point emphasized in this paper.

\subsection{Effect of Correlations}
\begin{figure}
\begin{center}
\includegraphics[width=0.52\textwidth,angle=0]{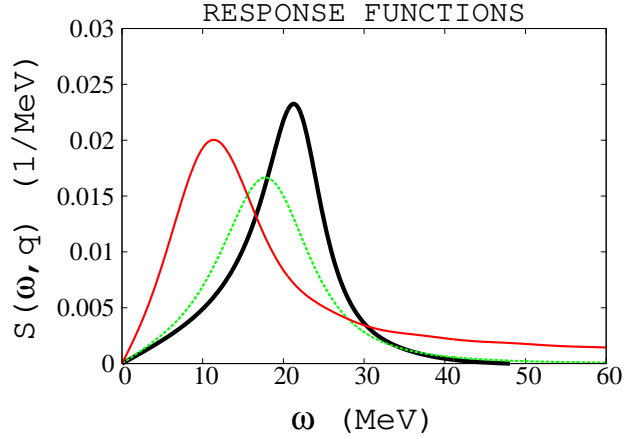}

\caption{
The full line to the right (black online) shows the response
function with  all self energies calculated with self-consistent 
correlated Green's functions. It relates to an
effective mass $m^{**}=0.61$. The left dotted line (red online)
shows the result with all selfenergies $\Sigma_{10}^{^{<}_{>}} \equiv 0$.
(In Fig. \ref{2Born}  denoted by $\Sigma_{10}$.)
The energy sum rule is here grossly violated.
The middle line (green online) neglects correlations between the
$G_{00}$ and $G_{01}$ Green's functions. (Last two diagrams in Fig.2) 
See text for further details.  All results are with $q_0=0.4$  fm$^{-1}$.
}
\label{full}
\end{center}
\end{figure}

The solid black (right) line in Fig. \ref{full} 
shows the response function calculated including all self energies i.e
with correlated Green's functions as shown by eqs (\ref{eq1}) and
(\ref{eq1a}). A comparison with the HF+RPA-result shown in Fig. \ref{HF} 
shows a considerable difference. Part of this difference is related to
the difference in effective masses $m^{**}$, 0.66 vs 0.61. 
Including only the second order self-energies $\Sigma_{00}^{^{<}_{>}}$
but neglecting the corresponding $\Sigma_{10}$ terms in eqs (\ref{S1})
,(\ref{S2}) and (\ref{S3}), (i.e. neglecting
the vertex-corrections) one finds a response-function as shown by the
left (red online) curve in Fig. \ref{full}. It shows the well-known
error in neglecting the vertex-corrections with a gross violation of the
sum-rule, eq. (\ref{sum}). Also shown (middle curve green online) is the
result when neglecting the contributions shown by eqs (\ref{S2}) and
(\ref{S3}).  (last two diagrams  in Fig. \ref{2Born}).

\begin{figure}
\begin{center}
\includegraphics[width=0.52\textwidth,angle=0]{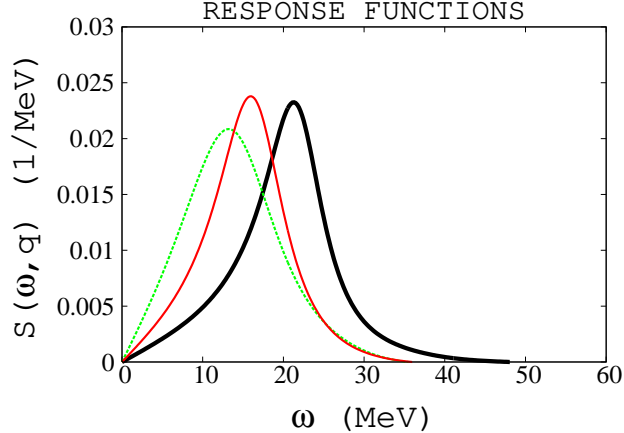}

\caption{
The full line to the right (black online) shows the response
function with  all self enrgies calculated with self-consistent 
correlated Green's functions.(Same as in Fig \ref{full}). 
The dotted line to the left (green online) shows the result
with $\Sigma^{HF}_{00}\equiv 0$ in the eqs for $G_{10}$ only. 
The effective mass is in this case 
$m^{**}=1$. The dotted line in the middle  (red online) was
obtained by assuming  $\Sigma^{HF}_{00}\equiv 0$ also in the eqs for $G_{00}$
, i.e.  also with $m^{**}=1$.  All results are with $q_0=0.4$ fm$^{-1}$.
}
\label{mass}
\end{center}
\end{figure}

The importance of the effective mass was already emphasized above in a
separate section..
A typical value for Brueckner and similar many-body calculations is
$m^*\sim 0.7$, consistent with the values shown above. There are however
the well-known corrections, (e.g. second and third order 'rearrangement'
corrections') that would bring this value up. Landau theory is more
compatible with an effective mass close to $m^*=1$.
We  therefore show in Fig \ref{mass} our result for this case. 
It is obtained by setting $\Sigma_{00}^{HF}=0$ with results shown in
Fig. \ref{mass}.  The sum-rule is consequently now satisfied with $m^{**}=1$.

\section{Summary and Conclusions}
A new computer program was designed to time-evolve 2-time Kadanoff-Baym
equations with self-energies computed with non-local separable
two-nucleon interactions. Previous program was restricted to the use of local
interactions only, while it is well known that a realistic representaion of
effective  nuclear forces are indeed non-local i.e. momentum dependent. A local
interaction is in momentum=space a function of momentum \it transfer \rm only.

The program was here used for the calculations of response functions for
symmetric nuclear matter. The all-important energy weighted sum-rule was found
to be well satisfied, validating the inegrity of the calculations.

Previous calculations presented in the literature are with few exceptions 
done in the HF+RPA approximation, i.e. neglecting the effect of correlations in
the nuclear medium. These correlations, related to the strong nuclear forces,
have been the focus of intense studies since the "birth" of nuclear physics.
It has been the purpose of this work to  investigate the effect of these
correlations on the  calculations of nuclear response. 
There are three separate (although related) effects to be expected. I. The
correlations result in a redistribution of occupied states as shown in Fig.
\ref{dens}. II. The selfenergis are complex. This causes a broadening of states
in general, while spectral functions of an uncorrelated medium are 
represented by delta-functions.  This broadening is of course the root of the
effect labelled by I. But it also causes a broadening of the response
functions as seen by comparing the full curves (black online) in Figs.
\ref{HF} and \ref{full}.
The third effect relates to the selfenergies $\Sigma_{10}$.
It is since a long time well known that the introduction of correlations in
numerical calculations as done here is not trivial. Selfenergy insertions in
propagators have to be accompanied by proper vertex corrections. This is
'automatically' accomplished  by the self-energies
denoted by $\Sigma_{10}$ in Fig. \ref{2Born}. Neglecting this term in the
calculations result in a gross violation of the sum rule as shown in Fig.
\ref{full}. 

An important factor to consider is also the effective mass. It has been 
a subject of numerous calculations and even more discussions in the literature.
(See section 3.3 above.)  In the context of response it is of course vital 
because it is as shown above, essential in determining the 'location' of 
the response along the $\omega$-axis.  It also affects the width of the
response function.
As was already discussed in  
section 3.3, there are numerous corrections that have to be included 
if a microscopic calculation is implemented.  Our calculations above yield 
a Brueckner (first order) estimate of $m^*$=0.6 to 0.7. Second and third 
order corrections may rise this to $m^*\sim 0.9$.  Empirical data (from 
experimental spectral densities) suggest a value close to
$m^*=1.0$.\cite{bro63} 

The effective mass is also an important factor as regards the
energy-weighted sum-rule. 
It was shown in an earlier work \cite{hsk17} that  if all selfenergies
were calculated consistently  with a \it local \rm interaction  the sum -rule is
satisfied with  $m^{**}=1$. (See above for definition of
$m^{**}$.)
If an external mean field $\Sigma_{00}$ was added the
sum-rule was then satisfied with the effective mass of this external
field. 

This situation is changed with the non-local interaction. The value of
$m^{**}$ was always found to be that of the chosen, not necessarily 
consistent, mean field $\Sigma_{00}$. It was however also illustrated
above that all selfenergies have to be included. The sum-rule would
otherwise be (sometimes grossly)  violated.

\newpage
\section{Acknowledgements}
I thank Dr George Papadimitriou for help with the nuclear matter
calculations and accompanying figures and Prof N.H. Kwong for numerous
discussions.
I thank The University of Arizona and in
particular the Department of Physics for providing office space and
access to computer facilities.

\bibliographystyle{apsrev4-1}

\end{document}